\input{aipcheck}

\documentclass[
    ,final            
  ]
  {aipproc}
\usepackage{aas_macros}
\layoutstyle{6x9}

\newcommand{\lsim}{\mathrel{\hbox{\rlap{\lower.55ex
\hbox{$\sim$}} \kern-.3em \raise.4ex \hbox{$<$}}}}

\begin{document}

\title{$^{30}$S Beam Development and X-ray Bursts}


\author{D. Kahl}{
  address={Center for Nuclear Study (CNS), the University of Tokyo, Wak\={o}, Saitama, 351-0198 Japan}
}

\author{A. A. Chen}{
  address={Department of Physics \& Astronomy, McMaster University, Hamilton, Ontario, L8S-4M1 Canada}
}

\author{S. Kubono}{
  address={Center for Nuclear Study (CNS), the University of Tokyo, Wak\={o}, Saitama, 351-0198 Japan}
}

\author{D. N. Binh}{
  address={Center for Nuclear Study (CNS), the University of Tokyo, Wak\={o}, Saitama, 351-0198 Japan}
}

\author{J. Chen}{
  address={Department of Physics \& Astronomy, McMaster University, Hamilton, Ontario, L8S-4M1 Canada}
}

\author{T. Hashimoto}{
  address={Center for Nuclear Study (CNS), the University of Tokyo, Wak\={o}, Saitama, 351-0198 Japan}
}

\author{S. Hayakawa}{
  address={Center for Nuclear Study (CNS), the University of Tokyo, Wak\={o}, Saitama, 351-0198 Japan}
}

\author{D. Kaji}{
  address={RIKEN (the Institute of Physical and Chemical Research), Wak\={o}, Saitama, 351-0198 Japan}
}

\author{A. Kim}{
  address={Department of Physics, Ewha Womans University, Seoul 120-750 Korea}
}

\author{Y. Kurihara}{
  address={Center for Nuclear Study (CNS), the University of Tokyo, Wak\={o}, Saitama, 351-0198 Japan}
}

\author{N. H. Lee}{
  address={Department of Physics, Ewha Womans University, Seoul 120-750 Korea}
}

\author{S. Nishimura}{
  address={RIKEN (the Institute of Physical and Chemical Research), Wak\={o}, Saitama, 351-0198 Japan}
}

\author{Y. Ohshiro}{
  address={Center for Nuclear Study (CNS), the University of Tokyo, Wak\={o}, Saitama, 351-0198 Japan}
}

\author{K. Setoodeh nia}{
  address={Department of Physics \& Astronomy, McMaster University, Hamilton, Ontario, L8S-4M1 Canada}
}

\author{Y. Wakabayashi}{
  address={Center for Nuclear Study (CNS), the University of Tokyo, Wak\={o}, Saitama, 351-0198 Japan}
  ,altaddress={Advanced Science Research Center, Japan Atomic Energy Agency (JAEA), Naka-gun, Ibaraki 319-1195, Japan}
}

\author{H. Yamaguchi}{
  address={Center for Nuclear Study (CNS), the University of Tokyo, Wak\={o}, Saitama, 351-0198 Japan}
}

\begin{abstract}
Over the past three years, we have worked on developing a well-characterized $^{30}$S radioactive beam to be used in a future experiment aiming to directly measure the $^{30}$S($\alpha$,p) stellar reaction rate within the Gamow window of Type I X-ray bursts.  
The importance of the $^{30}$S($\alpha$,p) reaction to X-ray bursts is discussed.  
Given the astrophysical motivation, the successful results of and challenges involved in the production of a low-energy $^{30}$S beam are detailed.  
Finally, an overview of our future plans regarding this on-going project are presented.
\end{abstract}

\maketitle

\section{Astrophysical Motivation}
The $^{30}$S($\alpha$,p) reaction is a significant link in the {\it $\alpha$p}-process, which competes with the {\it rp}-process in Type I X-ray bursts (XRBs) \cite{1981ApJS...45..389W}, but the reaction rate is virtually unconstrained by experimental data.
Not only are there no data directly measuring the $^{30}$S($\alpha$,p) cross section in the literature (at any energy), but there are no conclusive experimental reports on the nuclear structure of the compound nucleus $^{34}$Ar above the $\alpha$-threshold with regard to any possible $\alpha$-resonances.
Hydrodynamic models of XRBs indicate that variation of the theoretical reaction rate has significant consequences.

XRBs are understood to result from thermonuclear runaway in the hydrogen- and helium-rich accreted envelopes on the electron-degenerate surfaces of neutron star binary systems \cite{1976Natur.263..101W}.
Accretion ensures a steady flow of fresh material will spread around the surface of the neutron star only to be buried by the continuous pile-up of more matter. 
Equilibrium between a fierce gravitational pressure and the constant energy release of the $\beta$-limited CNO cycle is broken by a thin-shell instability, triggering the onset of explosive nucleosynthesis \cite{1974ApJ...191..479V, 1975ApJ...195..735H} .
Although powerful, these bursts do not disrupt the binary star system, hence X-ray bursters exhibit recurring episodes with hourly, daily, or more extended regularity, making them the ``most common thermonuclear explosions in the universe'' \cite{2008ApJS..174..261F}.
Computer simulations reproduce the energy release, burst profiles (rise time, peak wavelength, and decay curve), and the recurrence time-scales of these astrophysical phenomena.

The theoretical $^{30}$S($\alpha$,p) cross section at astrophysical energies is typically calculated using the statistical model of Hauser and Feshbach \cite{1952PhRv...87..366H}.
However, statistical models must assume an energy-dependent level density in the compound nucleus of interest, which is poorly constrained experimentally for the case of $^{34}$Ar.
In the case where a cross section is dominated by one or more narrow-resonant contribution(s), one ought to treat the problem via resonant reaction formalism and disfavor a statistical approach.
Previous work indicates that for $\alpha$-induced reactions on $T_{z}=\pm1$ ($T_{z}\equiv (N-Z)/2$) nuclei with $A=18$, the cross sections are shown to be dominated by natural-parity, $\alpha$-cluster resonances \cite{2002PhRvC..66e5802G, 2003PhRvC..68b5801D, 2003NuPhA.718..608B}.
These $\alpha$-capture trends are also observed at higher mass ($22 \leq A \leq 30$) on T$_{z}=1$ nuclei \cite{1999NuPhA.656....3A, 2005PrPNP..54..535A}, calling into question the accuracy of applying a statistical model to reaction rates such as $^{30}$S($\alpha$,p).

With nuclear reaction networks modeling the {\it rp}-process now extending beyond $A=100$ and involving a myriad of nuclear transmutations, one may wonder the extent to which variation of individual nuclear reaction rates within their uncertainties may give rise to gross systematic effects.
By varying only the $^{30}$S($\alpha$,p) reaction rate by a factor of 100, one model shows a different XRB profile, possibly accounting for the double-peaked structure in the bolometric luminosity of some rare systems \cite{2004ApJ...608L..61F}.
The $^{30}$S($\alpha$,p) reaction is also found to alter the crustal composition of neutron stars \cite{2006NuPhA.777..601S}, which influences the compositional inertia of recurrent XRBs \cite{1980ApJ...241..358T}.
A parametric study on the nuclear inputs to XRBs quantifies the effects of varying many reaction rates, indicating that $^{30}$S($\alpha$,p) is one of the nine most influential nuclear processes occurring below $^{56}$Ni and one of eight processes below $^{56}$Ni affecting the energy output by more than 5\% \cite{2008ApJS..178..110P}.
Clearly, experimental measurements to constrain the $^{30}$S($\alpha$,p) reaction rate are warranted.
\section{$^{30}$S Beam Production}
The physical chemistry of sulfur makes it extremely difficult to extract quickly and efficiently for re-acceleration as a secondary beam in ISOL facilities, and a method for producing low-energy ($<$ 4 MeV/u) secondary beams well-characterized in phase-space has yet to be realized using high-energy fragmentation techniques.  
As detailed presently, the low-energy Center for Nuclear Study (CNS) radioactive ion beam (CRIB) separator facility of the University of Tokyo \cite{2002EPJA...13..217K, 2005NIMPA.539...74Y} and located at the Nishina Center of RIKEN is capable of producing a $^{30}$S RI beam suitable for studying the astrophysical $^{30}$S($\alpha$,p) reaction.

\begin{figure}
  \includegraphics[clip=true, trim=0 0 0 420,height=.4\textheight]{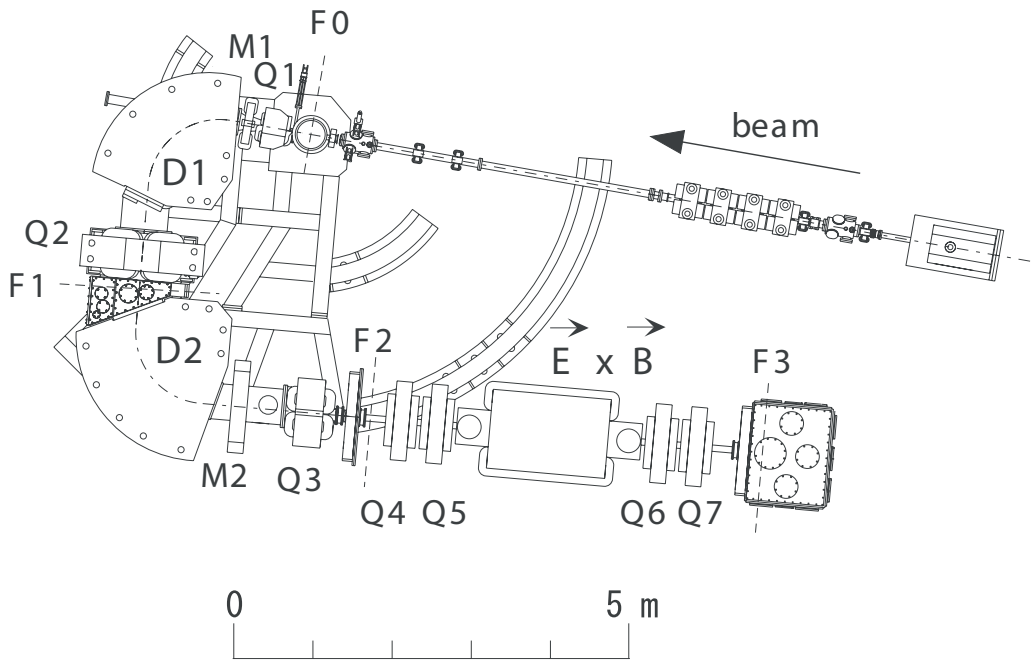}
\caption{Schematic overhead view of the CRIB separator facility.  The primary beam enters from the upper right side.  `F' stands for {\underline f}ocal plane, `Q' for magnetic {\underline q}uadruple, `M' for magnetic {\underline m}ultipole, `D' for magnetic {\underline d}ipole, and E $\times$ B is a Wien filter.  See the text.}
\label{crib_view}
\end{figure}
Heavy ion beams of a reasonably high intensity are extracted from the CNS 14 GHz HyperECR ion source \cite{RAPR.36.2003.279}, accelerated to an energy $\lsim 12$ MeV/u by the AVF cyclotron ($K=70$) at the RIKEN Accelerator Research Facility, and delivered to the entrance focal plane of CRIB (denoted `F0,' see below).
For the present work, we bombard a cryogenically-cooled $^{3}$He gas target with a stable $^{28}$Si beam, and via the $^{3}$He($^{28}$Si,$^{30}$S,)n reaction, we produce the RI species of interest.
The target $^{3}$He gas is confined by 2.5 $\mu$m Havar windows and force-flowed through a LN$_{2}$ cooling system to an effective target temperature around 80--90 K \cite{2008NIMPA.589..150Y}; the resulting $^{3}$He gas density is a factor of $\sim$3 greater than at ambient laboratory temperatures and stable against any density reduction effects induced by the energy deposition of the primary beam.

The cocktail beam emerging from the production target is mainly characterized and purified in the experiment hall by two magnetic dipoles and a Wien (velocity) filter, with beam-focusing magnetic multipoles surrounding these elements (see Figure \ref{crib_view}).
These beam-line components are separated by four focal planes of interest.
The primary beam focal point and the production target are located at F0, the dispersive focal plane between the two magnetic dipoles is denoted `F1,' the achromatic focal point after the second dipole `F2,' and the location of the experiment scattering chamber after the Wien filter `F3.'
Slits limiting the emittance and detectors for diagnosing the beam (such as silicon detectors and parallel plate avalanche counters (PPACs)) can be inserted or removed at all four and the latter three of the focal planes, respectively.
The slits located at F1 are of particular interest as they clearly define the magnetic rigidity ($B \rho$) of transmitted ions as well as their maximum momentum dispersion ($\Delta p/p$).

As we conducted $^{30}$S RI beam development tests in December 2006, May 2008, and July 2009 (two days each) varying many parameters to optimize results for $^{30}$S, we will limit the discussion to highlights of the most noteworthy points.
We tested three primary beams: $^{28}$Si$^{9+}$ of 6.9 MeV/u at 100 pnA, $^{28}$Si$^{10+}$ of 7.54 MeV/u at 10 pnA, and $^{28}$Si$^{9+}$ of 7.4 MeV/u at 144 pnA, listed in chronological test order.
All intensities quoted here were the maximum available at the time of the various tests, and due to beam-monitor counting limits at some focal planes, the actual intensity used may be lower where results are then normalized; previous work at CRIB indicates that, up to the rates of energy deposited in the production target for this work, linear beam-current normalization is applicable.

We found that $^{30}$S beam intensity shows a positive correlation with primary beam energy within this range, justifying our choice of the highest $^{28}$Si beam energy available from the cyclotron for each test; the results of recent improvements bringing the practical AVF cyclotron K-value closer to its design value are evident comparing the two $^{28}$Si$^{9+}$ beam energies listed above.
Although $\frac{d\sigma(E)}{dE}$ for $^{3}$He($^{28}$Si,$^{30}$S) is negative within this energy range \cite{1972NuPhA.198..449B, 1972PhRvC...6.1756G}, the combined energy dependence of the momentum straggling within the production target and the emerging charge-state distribution of $^{30}$S more than compensate for the decreasing $\sigma(E)$; RI beam production and ion-optical transport simulations are consistent with this interpretation of our experimental data.

The production target thickness was also optimized for $^{30}$S yield.
We experimentally found this value to be $\sim$1.7 mg/cm$^{2}$ of $^{3}$He, corresponding to a cryogenic gas pressure of 400 Torr.
We tested pressures of 200, 300 and 400 Torr for $^{30}$S production, and simulations as well as experimental tests with $^{3}$He($^{16}$O,$^{18}$Ne) indicate that significantly higher target thicknesses decreased RIB yield on-target.
These results can be understood in the same framework as the aforementioned primary beam energy effects: despite the linearly increasing isotropic yield of $^{30}$S for thicker targets, at some point the increasing momentum straggling and unfavorable shifting of the charge-state distribution from energy-loss considerations are dominate contributions to the overall intensity.

\begin{figure}
  \includegraphics[height=.4\textheight, angle=270]{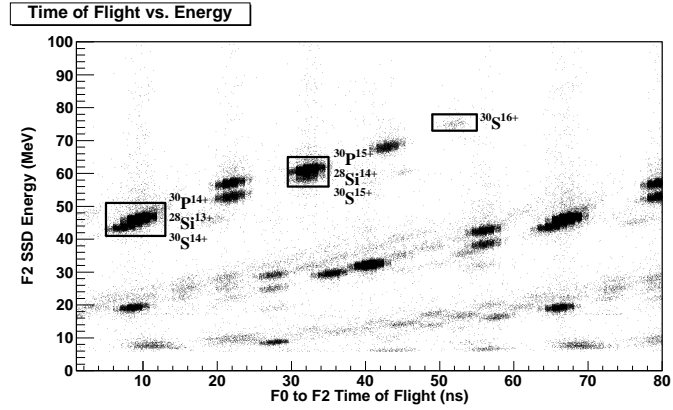}
\caption{The plot shows the particle flight time in nanoseconds on the abscissa and the residual energy in million electron volts on the ordinate for various nuclear species in the cocktail beam at the achromatic focal plane F2.  Although $^{30}$S$^{16+}$ is clearly separated, the loci of other charge-states of $^{30}$S are heavily contaminated.  The dispersive momentum slits are set such that $\Delta p/p \leq 0.625\%$.}
\label{pid_plot}
\end{figure}
Charge-state effects may seem unimportant since in principle we are free to tune the CRIB optics to any species of interest, and we may just choose the predominately populated ion-species of $^{30}$S.
To appreciate the importance of charge-state effects, one must consider that magnetic dipoles separate based on $p/q$, which, with a narrow momentum dispersion, effectively reduces to the charge-to-mass ratio $A/q$.
Owing to an extremely high intensity and statistical straggling effects, various charge states of the primary beam can be found at nearly any momenta.
Only the fully-stripped ion $^{30}$S$^{16+}$ ($A/q\sim1.875$) is clearly separated from the leaky primary beam (see Figure \ref{pid_plot}), which can never have $A/q<2$ ($^{28}$Si has $N=Z$ and hence $A=2Z$ and $q_{max}=Z$)\footnote{For simplicity, we quote $A/q$ without units.}.
$^{30}$P arising from the $^{3}$He($^{28}$Si,$^{30}$P)p reaction, with a cross section of the same order as the $^{3}$He($^{28}$Si,$^{30}$S) reaction \cite{1972NuPhA.198..449B}, also shows up as an impurity for all but the fully-stripped case of $^{30}$S.
In December 2006, we determined that without measuring the energy or a significant energy-loss of the cocktail beam, it was impossible to fully separate $^{30}$S$^{15+}$ from $^{28}$Si$^{14+}$ at a satisfactory level to avoid false-positives.
In May 2008 we could not purify $^{30}$S$^{+14}$ above the $\sim1\%$ level even with use of the Wien filter.
Although analysis of the July 2009 beam test is still underway, preliminary analysis indicates no significant improvement in the status of the $^{30}$S$^{14+}$ beam purity ($<2\%$).
Our results for $^{30}$S$^{16+}$ have continued to improve each year, and we have successfully achieved $\sim10^{4}$ particle Hz on target with $\sim25-30\%$ purity with $E_{beam}=30\pm3$ MeV.
Unfortunately at these energies, the $16+$ species of $^{30}$S is very weakly populated ($<1\%$), accounting for its low intensity.

As the production gas target has Havar windows, we considered ways of increasing the intensity of fully-stripped charge states of ions emerging from Havar foil.
We measured the charge-state distribution of $^{28}$Si beam ions in a thick carbon foil (550 $\mu$g/cm$^{2}$) compared to Havar foil (2.2 $\mu$m) (Table~\ref{tab1}); the energy-loss in Havar is slightly greater than carbon in this case, but the difference does not significantly effect our interpretation. 
For a $^{28}$Si beam of 3.4 MeV/u ($\sim E_{beam}$ of $^{30}$S), it was found that transmission of highly charged states of $^{28}$Si is improved through carbon foil compared to Havar foil with a ratio consistent with predictions of LISE++ (specifically, the {\it Global+Leon} model) \cite{2004NuPhA.746..411T, 2008NIMPB.266.4657T}.
In our July 2009 test, we used a 2.5 $\mu$m Be foil after the production target, which when normalized for comparison with the May 2008 results, indicates an increase in the $^{30}$S$^{16+}$ intensity by a factor of 2.
Although one theoretically expects this intensity increase to be on the order of a factor of 10--20, the Be foil was partially broken and perhaps did not cover the full solid-angle of beam emission from the production target, possibly accounting for this deficiency.
\begin{table}[h]
\centering
  \begin{tabular}{ c  c  c  }
      \hline
        {\normalsize Target} & {\normalsize Species} & {\normalsize Normalized} \\ 
	  & & pps @ 10 enA \\ \hline
	Havar & $^{28}$Si$^{12+}$ & $1.075 \times 10^{8}$ \\ 
	Havar & $^{28}$Si$^{13+}$ & $6.013 \times 10^{7}$ \\ 
	Havar & $^{28}$Si$^{14+}$ & $3.901 \times 10^{6}$ \\ 
	Carbon & $^{28}$Si$^{12+}$ & $1.758 \times 10^{8}$ \\ 
	Carbon & $^{28}$Si$^{13+}$ & $1.300 \times 10^{8}$ \\ 
	Carbon & $^{28}$Si$^{14+}$ & $4.365 \times 10^{7}$ \\ \hline
\end{tabular}
\caption
{Intensity of selected charge states of $^{28}$Si after passing through Havar foil or carbon foil.} 
\label{tab1}
\end{table}
\section{Planned Experimental Setup}
We plan to measure the $^{4}$He($^{30}$S,p) reaction\footnote{Although the experiment will be performed in inverse-kinematics, for nomenclatural convenience we may also write these reactions in normal kinematics ({\it e.g.} $^{30}$S($\alpha$,p)).} concurrently with $^{30}$S+$\alpha$ resonant elastic scattering using the thick target method \cite{1990SvJNP..52..408}.
The measurements will scan from the top of the 2 GK astrophysical Gamow window down to 1 GK, requiring a $^{30}$S $E_{beam}=32.3$ MeV on-target; while models indicate the explosive conditions in XRBs range from $0.4\leq T \leq 1.3$ GK \cite{2004ApJS..151...75W, 2008ApJS..174..261F}, the predicted cross sections at the lowest energies are much too low to be measured with the presently developed $^{30}$S beam.
Indeed, Hauser-Feshbach calculations indicate that we require a $^{30}$S beam with intensity 10$^{5}$ particle Hz to get reasonable ($\alpha$,p) reaction statistics in the 2 GK region over the course of an 11 day experiment.
But, with the presently-developed $^{30}$S beam at 10$^{4}$ particle Hz, we may reasonably measure the resonant $\alpha$ scattering of $^{30}$S and at least put a meaningful upper limit on the ($\alpha$,p) cross section; as the Fisker, Thielemann \& Wiescher model shows significant effects on the overall XRB light-curve for $^{30}$S($\alpha$,p) cross sections 10$^{2}$ above the Hauser-Feshbach rate \cite{2004ApJ...608L..61F}, we may at least confirm or rule-out such a rate with the planned experiment.

We will measure the reaction ejecta on an event-by-event basis in coincidence with $^{30}$S ions detected in two upstream beam monitors (PPACs or Multi-Channel Plates); as these beam monitors are limited to a counting rate of ~$\sim$10$^{6}$ particle Hz \cite{2001NIMPA.470..562K}, this puts an upper limit on the maximum beam intensity on-target.
The original experiment proposal called for a semi-cylindrical $^{4}$He gas cell and $\Delta$E-E silicon telescopes.
However, ejecta data collection in this setup is limited to a laboratory solid angle of $\sim$0.15 sr, and owing to the poor timing resolution of silicon detectors, clearly identifying ($\alpha$,p$_{0}$) events or distinguishing inelastic $\alpha$ events may be challenging.
Our group is currently developing an active-target method for helium gas, adding Gas Electron Multiplier (GEM) capabilities  to a newly designed Multi-Sampling and Tracking Proportional Counter (MSTPC) based on previous research \cite{2006NIMPA.556..339H}.
The active-target fill-gas is 90\% $^{4}$He and 10\% CO$_{2}$.
The GEM-MSTPC will allow for the full track-reconstruction of beam particles, reaction points, recoil nuclei, and ejecta, as well as full energy measurements of the ejecta in silicon detectors, allowing us to clearly identify reactions not transiting directly to ground-state levels.
The silicon detectors in the GEM-MSTPC cover $\sim$15\% of the laboratory solid angle, greatly increasing our data collection statistics compared to the originally proposed setup.
\section{Summary}
We successfully developed a $^{30}$S RI beam of 10$^{4}$ particle Hz of $\sim$25\% purity and $E_{beam}=30\pm 3$ MeV.
In the spring of 2010, we will conduct an experiment on the first measurement of the $^{30}$S+$^{4}$He system, the results of which will be applied to the astrophysical $^{30}$S($\alpha$,p) reaction rate.


\begin{theacknowledgments}
These experiments were made possible through the CNS and RIKEN collaboration.  The McMaster group is appreciative of funding from the National Science and Engineering Research Council of Canada.  The authors sincerely thank the Nishina Center beam operators.
\end{theacknowledgments}



\bibliographystyle{aipproc}   

\bibliography{/home/daid/library/library}

\IfFileExists{\jobname.bbl}{}
 {\typeout{}
  \typeout{******************************************}
  \typeout{** Please run "bibtex \jobname" to optain}
  \typeout{** the bibliography and then re-run LaTeX}
  \typeout{** twice to fix the references!}
  \typeout{******************************************}
  \typeout{}
 }

\end{document}